\documentclass{sig-alternate}

\usepackage[usenames]{color}
\usepackage{balance}
\usepackage{graphicx}
\usepackage{array}
\usepackage{colortbl}
\usepackage[protrusion=true,expansion]{microtype}
\usepackage{layout}
\usepackage{xspace}
\usepackage{multirow}
\usepackage{array}
\usepackage{colortbl}
\usepackage{enumitem}
\usepackage{url}
\usepackage{booktabs}
\usepackage{sfmath}
\usepackage{tabularx}
\usepackage{fixltx2e}
\usepackage{soul}
\sethlcolor{lightgray}

\usepackage{adjustbox}

\usepackage{rotating}

\usepackage{enumitem}
\setitemize{noitemsep,topsep=0pt,parsep=0pt,partopsep=0pt}

\usepackage{times}
\usepackage{cite}
\usepackage{graphicx}
\usepackage{epsfig}
\usepackage{xspace}
\usepackage{latexsym}
\usepackage[usenames]{color}
\usepackage{multirow}
\usepackage{listings}
\usepackage{textcomp}

\makeatletter
\newcommand\tabcaption{\def\@captype{table}\caption}
\makeatother

\newcommand{\ie}{\textit{i.e.,} }
\newcommand{\eg}{\textit{e.g.,} }

\newcommand\opt[1]{}
\newcommand\find[1]{}

\newcommand{\ls}[1]
   {\dimen0=\fontdimen6\the\font 
    \lineskip=#1\dimen0
    \advance\lineskip.5\fontdimen5\the\font
    \advance\lineskip-\dimen0
    \lineskiplimit=.9\lineskip
    \baselineskip=\lineskip
    \advance\baselineskip\dimen0
    \normallineskip\lineskip
    \normallineskiplimit\lineskiplimit
    \normalbaselineskip\baselineskip
    \ignorespaces
   }

\newenvironment{smalldescription}{
   \setlength{\topsep}{0pt}
   \setlength{\partopsep}{0pt}
   \setlength{\parskip}{0pt}
   \begin{description}
   \setlength{\leftmargin}{.2in}
   \setlength{\parsep}{0pt}
   \setlength{\parskip}{0pt}
   \setlength{\itemsep}{0pt}}{\end{description}}

\newcounter{observation}
\newcommand{\Y}{$\checkmark$}
\newcommand{\N}{$\times$}
\newcommand{\X}{$\otimes$}

\newcommand{\AEEE}{A$^3$E }
\newcommand{\ACTEVE}{ACTEve }

\begin{document}
%

\toappear

\title{Automated Test Input Generation for Android:\\Are We There Yet?}

\numberofauthors{3}
\author{
  \alignauthor Shauvik Roy Choudhary\\
  \affaddr{Georgia Institute of Technology, USA}\\
  \email{shauvik@cc.gatech.edu}
  \alignauthor Alessandra Gorla \\
  \affaddr{IMDEA Software Institute, Spain}\\
  \email{alessandra.gorla@imdea.org}
  \alignauthor Alessandro Orso \\
  \affaddr{Georgia Institute of Technology, USA} \\
  \email{orso@cc.gatech.edu}
}

\maketitle

\begin{abstract}
  Mobile applications, often simply called ``apps'', are increasingly
  widespread, and we use them daily to perform a number of activities.
  Like all software, apps must be adequately tested to gain confidence
  that they behave correctly. Therefore, in recent years, researchers
  and practitioners alike have begun to investigate ways to automate
  apps testing. In particular, because of Android's open source nature
  and its large share of the market, a great deal of research has been
  performed on input generation techniques for apps that run on the
  Android operating systems. At this point in time, there are in fact
  a number of such techniques in the literature, which differ in the
  way they generate inputs, the strategy they use to explore the
  behavior of the app under test, and the specific heuristics they
  use.  To better understand the strengths and weaknesses of these
  existing approaches, and get general insight on ways they could be
  made more effective, in this paper we perform a thorough comparison
  of the main existing test input generation tools for Android. In our
  comparison, we evaluate the effectiveness of these tools, and their
  corresponding techniques, according to four metrics: code coverage,
  ability to detect faults, ability to work on multiple platforms, and
  ease of use.  Our results provide a clear picture of the state of
  the art in input generation for Android apps and identify future
  research directions that, if suitably investigated, could lead to
  more effective and efficient testing tools for Android.
\end{abstract}

\section{Introduction}

In the past few years, we have witnessed an incredible boom of the
mobile applications (or simply, ``apps'') business. According to
recent reports, Google Play, the most popular app market for the
Android platform, currently offers over one million applications.
Similarly, the app market for Apple's iOS, the iTunes Store, provides
a comparable number of applications~\cite{appStoreNumbers}. The
prevalence of smartphones, tablets, and their applications is also
witnessed by the recent overtake of mobile apps over traditional
desktop applications in terms of Internet usage in the
US~\cite{mobileVsDesktop}.

Apps, like all software, must be adequately tested to gain confidence
that they behave correctly. It is therefore not surprising that, with
such a growth, the demand for tools for automatically testing mobile
applications has also grown, and with it the amount of research in
this area. Most of the researchers' and practitioners' efforts in this
area target the Android platform, for multiple reasons. First and
foremost, Android has the largest share of the mobile market at the
moment, which makes this platform extremely appealing for industry
practitioners. Second, due to the fragmentation of devices and OS
releases, Android apps often suffer from cross-platform and
cross-version incompatibilities, which makes manual testing of these
apps particularly expensive---and thus, particularly worth automating.
Third, the open-source nature of the Android platform and its related
technologies makes it a more suitable target for academic researchers,
who can get complete access to both the apps and the underlying
operating system. In addition, Android applications are developed in
Java.  Even if they are compiled into Dalvik bytecode, which
significantly differs from Java bytecode, there exist multiple
frameworks that can transform Dalvik bytecode into formats that are
more familiar and more amenable to analysis and instrumentation (\eg
Java Bytecode~\cite{Octeau:Dare:FSE:2012},
Jimple~\cite{Bartel:Dexpler:SOAP:2012}, and smali~\cite{smali}).

For all these reasons, there has been a great deal of research in
static analysis and testing of Android apps. In the area of testing,
in particular, researchers have developed techniques and tools to
target one of the most expensive software testing activities: test
input generation. There are in fact a number of these techniques in
the literature nowadays, which differ in the way they generate inputs,
the strategy they use to explore the behavior of the app under test,
and the specific heuristics they use. It is however still unclear what
are the strengths and weaknesses of these different approaches, how
effective they are in general and with respect to one another, and if
and how they could be improved.

To answer these questions, in this paper we present a comparative
study of the main existing test input generation techniques for
Android.\footnote{We had to exclude some techniques and tools from our
  study because either they were not available or we were not able to
  install them despite seeking their authors' help.} The goal of the
study is twofold. The first goal is to assess these techniques (and
corresponding tools) to understand how they compare to one another and
which ones may be more suitable in which context (\eg type of
apps). Our second goal is to get a better understanding of the general
tradeoffs involved in test input generation for Android and identify
ways in which existing techniques can be improved or new techniques be
defined.

In our comparison, we ran the tools considered on over 60 real-world
apps, while evaluating their effectiveness along several dimensions:
\emph{code coverage}, \emph{fault detection}, ability to work on
\emph{multiple platforms}, and \emph{ease of use}.
We considered coverage because test input generation tools should be
able to explore as much behavior of the application under test as
possible, and code coverage is typically used as a proxy for that. We
therefore used the tools to generate test inputs for each of the apps
considered and measured the coverage achieved by the different tools
on each app.
Although code coverage is a well understood and commonly used measure,
it is normally a gross approximation of behavior. Ultimately, test
input generation tools should generate inputs that are effective at
revealing faults in the code under test. For this reason, in our study
we also measured how many of the inputs generated by a tool resulted
in one or more failures (identified as uncaught exceptions) in the
apps considered.  We also performed additional manual and automated
checks to make sure that the thrown exceptions represented actual
failures.
Because of the fragmentation of the Android ecosystem, another
important characteristic for the tools considered is their ability to
work on different hardware and software configurations. We therefore
considered and assessed also this aspect of the tools, by running them
on different versions of the Android environment.
Finally, we also evaluated the ease of use of the tools, by assessing
how difficult it was to install and run them and the amount of manual
work involved in their use. Although this is a very practical aspect,
and one that normally receives only limited attention in research
prototypes, (reasonable) ease of use can enable replication studies
and allow other researchers to build on the existing technique and
tool.

Our results show that, although the existing techniques and tools we
studied are effective, they also have weaknesses and limitations, and
there is room for improvement. In our analysis of the results, we
discuss such limitations and identify future research directions that,
if suitably investigated, could lead to more effective and efficient
testing tools for Android. To allow other researchers to replicate our
studies and build on our work, we made all of our experimental
infrastructure and data publicly available at {\small
  \url{http://www.cc.gatech.edu/~orso/software/androtest}}.

The main contributions of this paper are:

\begin{itemize}
\item A survey of the main existing test input generation techniques
  for apps that run on the Android operating system.
\item An extensive comparative study of such techniques and tools
  performed on over 60 real-world Android apps.
\item An analysis of the results that discusses strengths and
  weaknesses of the different techniques considered and highlights
  possible future directions in the area.
\item A set of artifacts, consisting of experimental infrastructure as
  well as data, that are freely available and allow for replicating
  our work and building on it.
\end{itemize}

The remainder of the paper is structured as follows.
Section~\ref{sec:andr-platf} provides background information on the
Android environment and apps. Section~\ref{sec:overv-andr-test}
discusses the test input generation techniques and tools that we
consider in our study. Section~\ref{sec:evaluation} describes our
study setup and presents our results.  Section~\ref{sec:discussion}
analyzes and discusses our findings.  Finally,
Section~\ref{sec:conclusion} concludes the paper.

\section{The Android Platform}
\label{sec:andr-platf}

\begin{figure}[t]
  \centering
  \includegraphics[width=\linewidth]{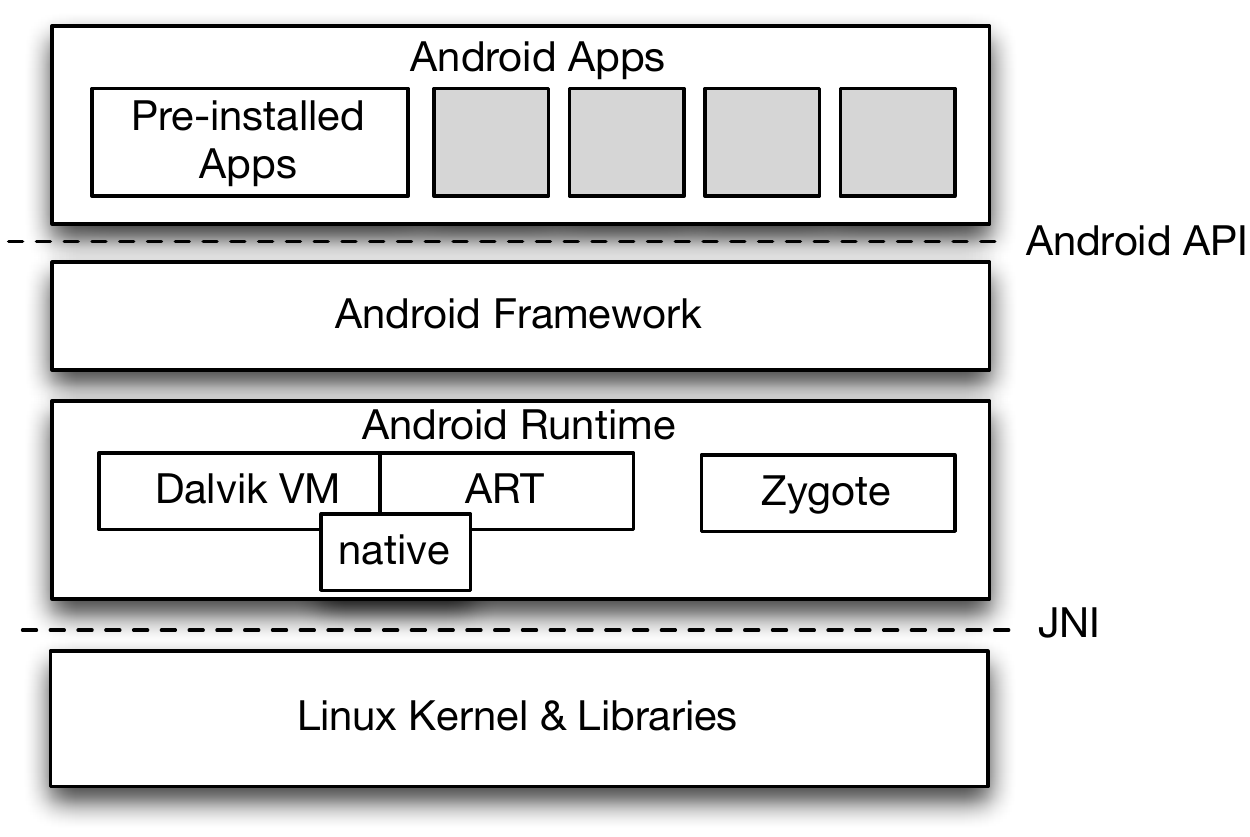}
  \caption{The Android architecture}
  \label{fig:android-architecture}
\end{figure}

Android applications are mainly written in Java, although some
high-performance demanding applications delegate critical parts of the
implementation to native code written in C or C++. During the build
process, Java source code gets first compiled into Java bytecode, then
translated into Dalvik bytecode, and finally stored into a machine
executable file in \textsf{.dex} format. Apps are distributed in the
form of \textsf{apk} files, which are compressed folders containing
\textsf{dex} files, native code (whenever present), and other
application resources.

Android applications run on top of a stack of three other main
software layers, as represented in
Figure~\ref{fig:android-architecture}.  The \emph{Android framework},
which lays below the Android apps, provides an API to access
facilities without dealing with the low level details of the operating
system. So far, there have been 20 different framework releases and
consequent changes in the API. Framework versioning is the first
element that causes the fragmentation problem in Android. Since it
takes several months for a new framework release to become predominant
on Android devices, most of the devices in the field run older
versions of the framework.  As a consequence, Android developers
should make an effort to make their apps compatible with older
versions of the framework, and it is therefore particularly desirable
to test apps on different hardware and software configurations before
releasing them.

In the \emph{Android Runtime} layer, the Zygote daemon manages the
applications' execution by creating a separate Dalvik Virtual Machine
for each running app. Dalvik Virtual Machines are register-based VMs
that can interpret the Dalvik bytecode. The most recent version of
Android includes radical changes in the runtime layer, as it
introduces ART (\ie Android Run-Time), a new runtime environment that
dramatically improves app performance and will eventually replace the
Dalvik VM.

At the bottom of the Android software stack stands a customized
\emph{Linux kernel}, which is responsible of the main functionality of
the system. A set of \emph{native code libraries}, such as WebKit,
libc and SSL, communicate directly with the kernel and provide a basic
hardware abstraction to the runtime layer.

\subsection*{Android applications}
\label{sec:android-applications}

Android applications declare their main components in the
\textsf{AndroidManifest.xml} file. Components can be of four different
types:

\begin{itemize}
\item \textsf{Activities} are the components in charge of an app's
  user interface. Each activity is a window containing various UI
  elements, such as buttons and text areas. Developers can control the
  behavior of each activity by implementing appropriate callbacks for
  each life-cycle phase (\ie created, paused, resumed, and destroyed).
  Activities react to user input events, such as clicks, and
  consequently are the primary target of testing tools for Android.

\item \textsf{Services} are application components that can perform
  long-running operations in the background. Unlike activities, they
  do not provide a user interface, and consequently they are usually
  not a direct target of Android testing tools, although they might be
  indirectly tested through some activities.

\item \textsf{Broadcast Receivers and Intents} allow inter-process
  communication. Applications can register broadcast receivers to be
  notified, by means of intents, about specific system events. Apps
  can thus, for instance, react whenever a new SMS is received, a new
  connection is available, or a new call is being made. Broadcast
  receivers can either be declared in the manifest file or at runtime,
  in the application code. In order to properly explore the behavior
  of an app, testing tools should be aware of what are the relevant
  broadcast receivers, so that they could trigger the right intents.

\item \textsf{Content Providers} act as a structured interface to
  shared data stores, such as contacts and calendar databases.
  Applications may have their own content providers and may make them
  available to other apps. Like all software, the behavior of an app
  may highly depend on the state of such content providers (\eg on
  whether the list of contacts is empty or it contains duplicates). As
  a consequence, testing tools should ``mock'' content providers in an
  attempt to make tests deterministic and achieve higher coverage of
  an app's behavior.
\end{itemize}

Despite being GUI-based and mainly written in Java, Android apps
significantly differ from Java standalone GUI applications and
manifest somehow different kinds of
bugs~\cite{Hu:GUITestingAndroid:AST:2011,Kechagia:AndroidFailures:EMSE:2014}. Existing
test input generation tools for
Java~\cite{Gross:Exsyst:ICSE:2012,Memon:GUITAR:WCRE:2003,Mariani:Autoblacktest:ICST:2012}
cannot therefore be straightforwardly used to test Android apps, and
custom tools must be adopted instead. For this reason, a great deal of
research has been performed in this area, and several test input
generation techniques and tools for Android applications have been
proposed. The next section provides an overview of the main existing
tools in this arena.


\section{Existing Android Testing Tools: an Overview}
\label{sec:overv-andr-test}

\newcolumntype{L}{>{\centering\arraybackslash}m{1.7cm}}
\newcolumntype{S}{>{\centering\arraybackslash}m{1cm}}
\newcolumntype{H}{>{\setbox0=\hbox\bgroup}c<{\egroup}@{}}

\begin{table*}
  \caption{Overview of existing test input generation tools for Android.}
  \centering 
  {\small
  \begin{tabular}[h]{|c|c|S|S|S|S|L|L|L|}
    \hline
    Name &
    Available &
    \multicolumn{2}{c|}{Instrumentation} &
    \multicolumn{2}{c|}{Events} &
    Exploration strategy &
    Needs source code &
    Testing strategy 
    \\ \cline{3-6}
    & & Platform & App & UI & System & & & \\
    \hline
    Monkey~\cite{monkey} & \Y & \N & \N & \Y &
    \N & Random & \N & Black-box\\
    Dynodroid~\cite{machiry13fse} & \Y & \Y & \N & \Y & \Y &
    Random & \N & Black-box \\

    DroidFuzzer~\cite{ye13momm} & \Y & \N & \N & \N & \N & Random &
    \N & Black-box\\
    IntentFuzzer~\cite{sasnauskas14woda} & \Y & \N & \N & \N & \N &
    Random & \N & White-box\\
    Null IntentFuzzer~\cite{nullfuzzer} & \Y & \N & \N & \N & \N &
    Random & \N & Black-box\\
    GUIRipper~\cite{amalfitano12ase} & \Y$^a$ & \N & \Y & \Y & \N &
    Model-based & \N & Black-box \\
    ORBIT~\cite{yang13fase} & \N & \N & \N & \Y & \N & Model-based
    & \Y  & Grey-box\\
    \AEEE-Depth-first~\cite{azim13oopsla} & \Y & \N & \Y & \Y & \N &
    Model-based & \N & Black-box\\
    SwiftHand~~\cite{choi13oopsla} & \Y & \N & \Y & \Y & \N &
    Model-based & \N & Black-box\\
    PUMA~\cite{Hao:PUMA:Mobisys:14} & \Y & \N & \Y & \Y & \N
    & Model-based & \N & Black-box \\
    \AEEE-Targeted~\cite{azim13oopsla} & \N & \N & \Y & \Y & \N &
    Systematic & \N & Grey-box\\
    EvoDroid~\cite{mahmood14fse} & \N & \N & \Y & \Y & \N &
    Systematic & \N & White-box\\
    ACTEve~\cite{anand12fse} & \Y & \Y & \Y & \Y
    & \Y & Systematic & \Y & White-box\\
    JPF-Android~\cite{vanderMerwe12sigsoft} & \Y & \N & \N & \Y
    & \N & Systematic & \Y & White-box\\
    \hline
  \end{tabular}}

  \vspace{.05cm}
  \raggedright{a) Not open source.}
  \label{tab:overview-tools}
\end{table*}

As we mentioned in the Introduction, there are several existing test
input generation tools for Android.  The primary goal of these tools
is to detect existing faults in Android apps, and app developers are
thus typically the main stakeholders, as they can automatically test
their application and fix discovered issues before deploying it.  The
dynamic traces generated by these tools, however, can also be the
starting point of more specific analyses, which can be of primary
interest of Android market maintainers and final users. In fact,
Android apps heavily use features such as native code, reflection and
code obfuscation that hit the limitations of almost every static
analysis tool~\cite{FlowDroid,Gorla:CHABADA:ICSE:2014}.  Thus, to
explore the behavior of Android apps and overcome such limitations it
is common practice to resort to dynamic analysis and use test input
generation tools to explore enough behaviors for the analysis.
Google, for instance, is known to run every app on its cloud
infrastructure to simulate how it might work on user devices and look
for malicious behavior~\cite{google-bouncer}. Only apps that pass this
test are listed in the Google Play market.  Finally, users can analyze
apps focusing on specific aspects, such as observing possible leaks of
sensitive information~\cite{Enck:Taintdroid:OSDI:2010} and profiling
battery, memory, or networking
usage~\cite{Wei:ProfileDroid:mobicom:2012}.

Test input generation tools can either analyze the app in isolation or
focus on the interaction of the app with other apps or the underlying
framework.  Whatever is the final usage of these tools, the challenge
is to generate relevant inputs to exercise as much behavior of
application under test as possible.

As Android apps are event-driven, inputs are normally in the form of
events, which can either mimic user interactions (\emph{UI events}),
such as clicks, scrolls, and text inputs, or \emph{system events},
such as the notification of a newly received SMS. Testing tools can
generate such inputs following different strategies. They can generate
them \emph{randomly} or follow a \emph{systematic} exploration
strategy.  In this latter case, exploration can either be guided by a
\emph{model} of the app, which constructed statically or dynamically,
or exploit techniques that aim to achieve as much code coverage as
possible. Along a different dimension, testing tools can generate
events by considering Android apps as either a \emph{black box} or a
\emph{white box}. In this latter case, they would consider the code
structure. \emph{Grey box} approaches are also possible, which
typically extract high-level properties of the app, such as the list
of activities and the list of UI elements contained in each activity,
in order to generate events that will likely expose unexplored
behavior.

Table~\ref{tab:overview-tools} provides an overview of the test input
generation tools for Android that have been presented in different
venues. To the best of our knowledge this list is complete. The table
reports all these tools, and classifies them according to the metrics
reported above. Moreover, the tables reports other relevant features
of each tool, such as whether it is \emph{publicly available} or
rather it is only described in a paper or its distribution is under
restricted policies of a company, whether the tool \emph{requires the
  source code} of the application under test, and whether it requires
\emph{instrumentation}, either at the application level or of the
underlying Android framework. The following sections report more
details of each of these tools.


\subsection{Random exploration strategy}
\label{sec:rand-expl-strat}

The first class of test input generation tools we consider employs a
random strategy to generate inputs for Android apps. In the most
simple form, the random strategy generates only UI events. Randomly
generating system events would be highly inefficient, as there are too
many such events, and applications usually react to only few of them,
and only under specific conditions.

Many tools that fall in this category aim to test inter-application
communications by randomly generating values for
\texttt{Intents}. Intent fuzzers, despite being test input generators,
have quite a different purpose. By randomly generating inputs, they
mainly generate invalid ones, thus testing the robustness of an app.
These tools are also quite effective at revealing security
vulnerabilities, such as denial-of-service vulnerabilities. We now
provide a brief description of the tools that fall in this category
and their salient characteristics.

\begin{description}
\addtolength{\itemsep}{-2mm}
\item[Monkey~\cite{monkey}] is the most frequently used tool to test
  Android apps, since it is part of the Android developers toolkit and
  thus does not require any additional installation effort. Monkey
  implements the most basic random strategy, as it considers the app
  under test a black-box and can only generate UI events. Users have
  to specify the number of events they want Monkey to generate. Once
  this upper bound has been reached, Monkey stops.

\item[Dynodroid~\cite{machiry13fse}] is also based on random
  exploration, but it has several features that make its exploration
  more efficient compared to Monkey. First of all, it can generate
  system events, and it does so by checking which ones are relevant
  for the application. Dynodroid gets this information by monitoring
  when an application registers a listener within the Android
  framework. For this reason it requires to instrument the framework.

  The random event generation strategy of Dynodroid is smarter than
  the one that Monkey implements. It can either select the events that
  have been least frequently selected (\textsf{Frequency} strategy)
  and can keep into account the context (\textsf{BiasedRandom}
  strategy), that is, events that are relevant in more contexts will
  be selected more often. For our study we used the
  \textsf{BiasedRandom} strategy, which is the default one.

  An additional improvement of Dynodroid is the ability to let users
  manually provide inputs (\eg for authentication) when the
  exploration is stalling.

\item[Null intent fuzzer~\cite{nullfuzzer}] is an open-source basic
  intent fuzzer that aims to reveal crashes of activities that do not
  properly check input intents. While quite effective at revealing
  this type of problems, it is fairly specialized and not effective at
  detecting other issues.

\item[Intent Fuzzer~\cite{sasnauskas14woda}] mainly tests how an app
  can interact with other apps installed on the same device. It
  includes a static analysis component, which is built on top of
  FlowDroid~\cite{FlowDroid}, for identifying the expected structure
  of intents, so that the fuzzer can generate them accordingly. This
  tool has shown to be effective at revealing security issues. Maji et
  al.\ worked on a similar intent
  fuzzer~\cite{Maji:robustness:DSN:2012}, but their tool has more
  limitations than Intent Fuzzer.


\item[DroidFuzzer~\cite{ye13momm}] is different from other tools that
  mainly generate UI events or intents. It solely generates inputs for
  activities that accept MIME data types such as AVI, MP3, and HTML
  files. The authors of the paper show how this tool could make some
  video player apps crash. DroidFuzzer is supposed to be implemented
  as an Android application. However, it is not available, and the
  authors did not reply to our request for the tool.
\end{description}

In general, the advantage of random test input generators is that they
can efficiently generate events, and this makes them particularly
suitable for stress testing. However, a random strategy would hardly
be able to generate highly specific inputs. Moreover, these tools are
not aware of how much behavior of the application has been already
covered, and thus are likely to generate redundant events that do not
help the exploration. Finally, they do not have a stopping criterion
that indicates the success of the exploration, but rather resort to a
manually specified timeout.

\subsection{Model-based exploration strategy}
\label{sec:model-based-expl}

Following the example of several Web
crawlers~\cite{crawljax:tweb12,webmate-swqd-2013,RoyChoudhary:XPERT:ICSE:2013}
and GUI testing tools for stand alone
applications~\cite{Gross:Exsyst:ICSE:2012,Mariani:Autoblacktest:ICST:2012,Memon:GUITAR:WCRE:2003},
some Android testing tools build and use a GUI model of the
application to generate events and systematically explore the behavior
of the application. These models are usually finite state machines
that have activities as states and events as transitions. Some tools
build more precise models by differentiating the state of activity
elements when representing states (\eg the same activity with a button
enabled and disabled would be represented as two separate states).
Most tools build such model dynamically and terminate when all the
events that can be triggered from all the discovered states lead to
already explored states.

\begin{description}
\addtolength{\itemsep}{-2mm}
\item[GUIRipper~\cite{amalfitano12ase}], which later became
  MobiGUITAR~\cite{MobiGUITARIEEESoftware2014}, dynamically builds a
  model of the app under test by crawling it from a starting state.
  When visiting a new state, it keeps a list of events that can be
  generated on the current state of the activity, and systematically
  triggers them. GUIRipper implements a DFS strategy, and it restarts
  the exploration from the starting state when it cannot detect new
  states during the exploration. It generates only UI events, thus it
  cannot expose behavior of the app that depend on system events.
  GUIRipper has two characteristics that make it unique among
  model-based tools. First, it allows for exploring an application
  from different starting states. (This, however, has to be done
  manually.) Moreover, it allows testers to provide a set of input
  values that can be used during the exploration.  GUIRipper is
  publicly available, but unfortunately it is not open source, and its
  binary version is compiled for Windows machines.

\item[ORBIT~\cite{yang13fase}] implements the same exploration
  strategy of GUIRipper, but statically analyzes the application's
  source code to understand which UI events are relevant for a
  specific activity. It is thus supposed to be more efficient than
  GUIRipper, as it should generate only relevant inputs. However, the
  tool is unfortunately not available, as it is propriety of Fujitsu
  Labs.  It is unclear whether ORBIT requires any instrumentation of
  the platform or of the application to run, but we believe that this
  is not the case.

\item[\AEEE-Depth-first~\cite{azim13oopsla}] is an open source tool
  that implements two totally distinct and complementary strategies.
  The first one implements a depth first search on the dynamic model
  of the application. In essence, it implements the exact same
  exploration strategy of the previous tools. Its model representation
  is more abstract than the one used by other tools, as it represents
  each activity as a single state, without considering different
  states of the elements of the activity. This abstraction does not
  allow the tool to distinguish some states that are different, and
  may lead to missing some behavior that would be easy to exercise if
  a more accurate model where to be used.

\item[SwiftHand~\cite{choi13oopsla}] has, as its main goal, that to
  maximize the coverage of the application under test. Similarly to
  the previously mentioned tools, SwiftHand uses a dynamic finite
  state machine model of the app, and one of its main characteristics
  is that it optimizes the exploration strategy to minimize the
  restarts of the app while crawling. SwiftHand generates only
  touching and scrolling UI events and cannot generate system events.

\item[PUMA~\cite{Hao:PUMA:Mobisys:14}] is a novel tool that includes a
  generic UI automator that provides the random exploration also
  implemented by Monkey. The novelty of this tool is not in the
  exploration strategy, but rather in its design. PUMA is a framework
  that can be easily extended to implement any dynamic analysis on
  Android apps using the basic monkey exploration strategy. Moreover,
  it allows for easily implementing different exploration strategies,
  as the framework provides a finite state machine representation of
  the app. It also allows for easily redefining the state
  representation and the logic to generate events. PUMA is publicly
  available and open source. It is, however, only compatible with the
  most recent releases of the Android framework.

\end{description}

Using a model of the application should intuitively lead to more
effective results in terms of code coverage. Using a model would, in
fact, limit the number of redundant inputs that a random approach
generates.
The main limitation of these tools stands in the state representation
they use, as they all represent new states only when some event
triggers changes in the GUI. Some events, however, may change the
internal state of the app without affecting the GUI. In such
situations these algorithm would miss the change, consider the event
irrelevant, and continue the exploration in a different direction. A
common scenario in which this problem occurs is in the presence of
services, as services do not have any user interface (see
Section~\ref{sec:android-applications}).

\subsection{Systematic exploration strategy}
\label{sec:syst-expl-strat}

Some application behavior can only be revealed upon providing specific
inputs. This is the reason why some Android testing tools use more
sophisticated techniques such as symbolic execution and evolutionary
algorithms to guide the exploration towards previously uncovered code.

\begin{description}
\addtolength{\itemsep}{-2mm}
\item[\AEEE-Targeted~\cite{azim13oopsla}] provides an alternative
  exploration strategy that complements the one described in
  Section~\ref{sec:model-based-expl}. The targeted approach relies on
  a component that, by means of taint analysis, can build the Static
  Activity Transition Graph of the app. Such graph is an alternative
  to the dynamic finite state machine model of the depth-first search
  exploration and allows the tool to cover activities more efficiently
  by generating intents. While the tool is available on a public
  repository, this strategy does not seems to be.

\item[EvoDroid~\cite{mahmood14fse}] relies on evolutionary algorithms
  to generate relevant inputs. In the evolutionary algorithms
  framework, EvoDroid represents individuals as sequences of test
  inputs and implements the fitness function so as to maximize
  coverage.  EvoDroid is no longer publicly available. In
  Section~\ref{sec:selected-tools}, we provide more details about
  this.

\item[\ACTEVE~\cite{anand12fse}] is a concolic-testing tool that
  symbolically tracks events from the point in the framework where
  they are generated up to the point where they are handled in the
  app. For this reasons, \ACTEVE needs to instrument both the
  framework and the app under test. \ACTEVE handles both system and UI
  events.

\item[JPF-Android~\cite{vanderMerwe14sigsoft}] extends Java PathFinder
  (JPF), a popular model checking tool for Java, to support Android
  apps. This would allow to verify apps against specific properties.
  Liu et al.\ were the first who investigated the possibility of
  extending JPF to work with Android apps~\cite{LiuJPF-Android}. 
  What they present, however, is mainly a feasibility study. They
  themselves admit that developing the necessary components would
  require a lot of additional engineering efforts.  Van Der Merwe et
  al.\ went beyond that and properly implemented and open sourced the
  necessary extensions to use JPF with Android. JPF-Android aim to
  explore all paths in an Android app and can identify deadlocks and
  runtime exceptions.  The current limitations of the tool, however,
  seriously limit its practical applicability.
\end{description}

Implementing a systematic strategy leads to clear benefits in
exploring behavior that would be hard to reach with random techniques.
Compared to random techniques, however, these tools are considerably
less scalable.


\section{Empirical Study}
\label{sec:evaluation}

To evaluate the test input generation tools that we considered (see
Section~\ref{sec:overv-andr-test}), we deployed them along with a
group of Android applications on a common virtualized
infrastructure. Such infrastructure aims to ease the comparison of
testing tools for Android, and we make it available such that
researchers and practitioners can easily evaluate new Android testing
tools against existing ones in the future. Our evaluation does not
include all the tools listed in Table~\ref{tab:overview-tools}. We
explain the reasons why we had to exclude some of the tools in
Section~\ref{sec:selected-tools}. The following sections provide more
details on the virtualized infrastructure (Section~\ref{sec:protocol})
and on the set of mobile apps that we considered as benchmarks for our
study (Section~\ref{sec:benchmarks}).

Our study evaluated these tools according to four main criteria:

\begin{description}
\addtolength{\itemsep}{-2mm}

\item[C1: Effectiveness of the exploration strategy.] The inputs that
  \- these tools generate should ideally cover as much behavior as
  possible of the app under test. Since \emph{code coverage} is a
  common proxy to estimate behavior coverage, we evaluate the
  statement coverage that each tool achieves on each benchmark. We also
  report a comparison study, in terms of code coverage, among
  different tools.

\item[C2: Fault detection ability.] The primary goal of test input
  generators is to expose existing faults. We therefore evaluate, for
  each tool, how many failures it triggers for each app, and we then
  compare the effectiveness of different tools in terms of the
  failures they trigger.

\item[C3: Ease of use.] Usability should be a primary concern for all
  tool developers, even when tools are just research prototypes. We
  evaluate the usability of each tool by considering how much effort
  it took us to install and use it.

\item[C4: Android Framework Compatibility.] One of the major problems
  in the Android ecosystem is fragmentation. Test input generation
  tools for Android should therefore ideally run on multiple versions
  of the Android framework, so that developers could assess how their
  app behaves in different environments.

\end{description}

Each of these research questions is addressed separately in
Section~\ref{sec:rq1} (C1), Section~\ref{sec:rq2} (C2),
Section~\ref{sec:rq2} (C2), and Section~\ref{sec:rq4} (C4).

\subsection{Selected Tools}
\label{sec:selected-tools}

Our evaluation could not consider all the tools that we list in
Table~\ref{tab:overview-tools}. First, we decided to ignore intent
fuzzers, as these tools do not aim to test the whole behavior of an
app, but rather to expose possible security vulnerabilities.
Therefore, it would have been hard to compare the results provided by
these tools with other test input generators. We initially considered
the possibility of evaluating DroidFuzzer, IntentFuzzer, and Null
Intent Fuzzer separately. However, Null Intent Fuzzer requires to
manually select each activity that the fuzzer should target, and it is
therefore not feasible to evaluate it on large scale experiments.
Moreover, we had to exclude DroidFuzzer because the tool is not
publicly available, and we were not successful in reaching the authors
by email.

Among the rest of the tools, we had to exclude also EvoDroid and
ORBIT.  EvoDroid used to be publicly available on its project website,
and we tried to install and run it. We also contacted the authors
after we ran into some problems with missing dependencies, but despite
their willingness to help, we never managed to get all the files we
needed and the tool to work. Obtaining the source code and fixing the
issues ourselves was unfortunately not an option, due to the
contractual agreements with their funding agencies. Moreover, at the
time of writing the tool is no longer available, even as a
closed-source package. The problem with ORBIT is that it is a
proprietary software of Fujitsu Labs, and therefore the authors could
not share the tool with us.

We also excluded JPF-Android, even if the tool is publicly
available. This tool in fact, expects users to manually specify the
input sequence that JPF should consider for verification, and this
would have been time consuming to do for all the benchmarks we
considered.

Finally, we could not evaluate the \AEEE-targeted, since this
strategy was not available in the public distribution of the
tool at the time of our experiments.

\subsection{Mobile App Benchmarks}
\label{sec:benchmarks}

To evaluate the selected tools, we needed a common set of
benchmarks. Obtaining large sets of Android apps is not an issue, as
binaries can be directly downloaded from app markets, and there are
many platforms such as F-Droid~\footnote{\url{http://f-droid.org}}
that distribute open-source Android applications. However, since many
of these tools are not maintained, and therefore could easily crash on
apps that utilize recent features. Thus, for our experiments we
combined all the open source mobile application benchmarks that were
considered in the evaluation of at least one tool. We retrieved the
same version of the benchmarks, as they were reported in each paper.

PUMA and \AEEE were originally evaluated on a set of apps downloaded
from the Google Play market. We excluded these apps from our dataset
because some tools need the application source code, and therefore it
would have been impossible to run them on these benchmarks.

We collected \emph{68 applications} in total. 52 of them come from the
Dynodroid paper~\cite{machiry13fse}, 3 from
GUIRipper~\cite{amalfitano12ase}, 5 from \ACTEVE~\cite{anand12fse},
and 10 from SwiftHand~\cite{choi13oopsla}.  Table~\ref{tableApps}
reports the whole list of apps that we collected, together with the
corresponding version and category. For each app we report whether it
was part of the evaluation benchmarks of a specific tool, and we also
report whether during our evaluation the tools failed in analyzing it.

\begin{table}[!htb]
  \caption{List of subject apps. (\Y~indicates that the app was used originally in the tool's evaluation and \X~indicates that the app crashed while being exercised by the tool in our experiments)}
  \scriptsize
  \bgroup
  \setlength{\tabcolsep}{.4em}
  \begin{tabularx}{\linewidth}{|l|r|l|c|p{0.1cm}p{0.1cm}|p{0.1cm}p{0.1cm}|p{0.1cm}p{0.1cm}|p{0.1cm}p{0.1cm}|p{0.1cm}p{0.1cm}|p{0.1cm}p{0.1cm}|}
    \hline
    \multicolumn{3}{|c|}{\textbf{Subject}}  & \begin{turn}{90}\multirow{2}{*}{\textbf{Monkey}}\end{turn}   & \multicolumn{2}{l|}{ \begin{turn}{90}\textbf{\ACTEVE}\end{turn} } & \multicolumn{2}{l|}{\begin{turn}{90}\textbf{DynoDroid}\end{turn} } & \multicolumn{2}{l|}{\begin{turn}{90}\textbf{\AEEE}\end{turn} } & \multicolumn{2}{l|}{\begin{turn}{90}\textbf{GuiRipper}\end{turn} } & \multicolumn{2}{l|}{\begin{turn}{90}\textbf{PUMA}\end{turn} } & \multicolumn{2}{l|}{\begin{turn}{90}\textbf{SwiftHand}\end{turn} } \\ \cline{1-3}
    {Name}     &  {Ver.}  &    {Category}   &         &        &          & &          &       &       &          &          &        &       &          &          \\ \hline
    Amazed               &2.0.2             &Casual              & \X               &                   &                  & \Y                  &                    &                 &                 &                    & \X                  &                  &                 &                    &                    \\
    AnyCut               &0.5               &Productiv.        &                 &                   &                  & \Y                  &                    &                 &                 &                    &                    &                  &                 &                    &                    \\
    Divide\&Conquer      &1.4               &Casual              &                 &                   &                  & \Y                  &                    &                 &                 &                    & \X                  &                  &                 &                    &                    \\
    LolcatBuilder        &2                 &Entertain.               &                 &                   &                  & \Y                  &                    &                 &                 &                    &                    &                  &                 &                    &                    \\
    MunchLife            &1.4.2             &Entertain.               &                 &                   &                  & \Y                  &                    &                 &                 &                    &                    &                  &                 &                    &                    \\
    PasswordMakerPro     &1.1.7             &Tools               &                 &                   &                  & \Y                  &                    &                 &                 &                    &                    &                  &                 &                    &                    \\
    Photostream          &1.1               &Media               &                 &                   &                  & \Y                  &                    &                 &                 &                    &                    &                  &                 &                    &                    \\
    QuickSettings        &1.9.9.3           &Tools               &                 &                   &                  & \Y                  &                    &                 &                 &                    &                    &                  &                 &                    &                    \\
    RandomMusicPlay      &1                 &Music               &                 & \Y                &                  & \Y                  &                    &                 &                 &                    &                    &                  &                 &                    &                    \\
    SpriteText           &-                 &Sample              &                 &                   &                  & \Y                  &                    &                 &                 &                    &                    &                  &                 &                    &                    \\
    SyncMyPix            &0.15              &Media               &                 &                   &                  & \Y                  &                    &                 &                 &                    &                    &                  &                 &                    &                    \\
    Triangle             &-                 &Sample              &                 &                   &                  & \Y                  &                    &                 &                 &                    &                    &                  &                 &                    &                    \\
    A2DP Volume          &2.8.11            &Transport      &                 &                   &                  & \Y                  &                    &                 &                 &                    &                    &                  &                 &                    &                    \\
    aLogCat              &2.6.1             &Tools               &                 &                   &                  & \Y                  &                    &                 &                 &                    &                    &                  &                 &                    &                    \\
    AardDictionary       &1.4.1             &Reference           &                 &                   &                  & \Y                  &                    &                 &                 &                    &                    &                  &                 &                    &                    \\
    BaterryDog           &0.1.1             &Tools               &                 &                   &                  & \Y                  &                    &                 &                 &                    &                    &                  &                 &                    &                    \\
    FTP Server           &2.2               &Tools               &                 &                   &                  & \Y                  &                    &                 &                 &                    &                    &                  &                 &                    &                    \\
    Bites                &1.3               &Lifestyle           &                 &                   &                  & \Y                  &                    &                 &                 &                    &                    &                  &                 &                    &                    \\
    Battery Circle       &1.81              &Tools               &                 &                   &                  & \Y                  &                    &                 &                 &                    &                    &                  &                 &                    &                    \\
    Addi                 &1.98              &Tools               &                 &                   &                  & \Y                  &                    &                 &                 &                    &                    &                  &                 &                    &                    \\
    Manpages             &1.7               &Tools               &                 &                   &                  & \Y                  &                    &                 &                 &                    &                    &                  &                 &                    &                    \\
    Alarm Clock          &1.51              &Productiv.        &                 &                   &                  & \Y                  &                    &                 &                 &                    &                    &                  &                 &                    &                    \\
    Auto Answer          &1.5               &Tools               &                 &                   &                  & \Y                  &                    &                 &                 &                    &                    &                  &                 &                    &                    \\
    HNDroid              &0.2.1             &News                &                 &                   &                  & \Y                  &                    &                 &                 &                    &                    &                  &                 &                    &                    \\
    Multi SMS            &2.3               &Comm.       &                 &                   &                  & \Y                  &                    &                 &                 &                    &                    &                  &                 &                    &                    \\
    World Clock          &0.6               &Tools               &                 &                   &                  & \Y                  &                    &                 &                 &                    &                    &                  &                 &                    & \X                  \\
    Nectroid             &1.2.4             &Media               &                 &                   &                  & \Y                  &                    &                 &                 &                    &                    &                  &                 &                    &                    \\
    aCal                 &1.6               &Productiv.        &                 &                   &                  & \Y                  &                    &                 &                 &                    &                    &                  &                 &                    &                    \\
    Jamendo              &1.0.6             &Music               &                 &                   &                  & \Y                  &                    &                 &                 &                    &                    &                  &                 &                    &                    \\
    AndroidomaticK.   &1.0                &Comm.       &                 &                   & \X                & \Y                  &                    &                 &                 &                    &                    &                  &                 &                    &                    \\
    Yahtzee              &1                 &Casual              &                 &                   &                  & \Y                  &                    &                 &                 &                    &                    &                  &                 &                    &                    \\
    aagtl                &1.0.31            &Tools               &                 &                   &                  & \Y                  &                    &                 &                 &                    &                    &                  &                 &                    &                    \\
    Mirrored             &0.2.3             &News                &                 &                   &                  & \Y                  &                    &                 &                 &                    &                    &                  &                 &                    &                    \\
    Dialer2              &2.9               &Productiv.        &                 &                   &                  & \Y                  &                    &                 &                 &                    &                    &                  &                 &                    &                    \\
    FileExplorer         &1                 &Productiv.        &                 &                   &                  & \Y                  &                    &                 &                 &                    &                    &                  &                 &                    &                    \\
    Gestures             &1                 &Sample              &                 &                   &                  & \Y                  &                    &                 &                 &                    &                    &                  &                 &                    &                    \\
    HotDeath             &1.0.7             &Card                &                 &                   &                  & \Y                  &                    &                 &                 &                    &                    &                  &                 &                    &                    \\
    ADSdroid             &1.2               &Reference           & \X               &                   &                  & \Y                  &                    &                 &                 &                    &                    &                  &                 &                    &                    \\
    myLock               &42                &Tools               &                 &                   &                  & \Y                  &                    &                 &                 &                    &                    &                  &                 &                    &                    \\
    LockPatternGen.      &2                 &Tools               &                 &                   &                  & \Y                  &                    &                 &                 &                    &                    &                  &                 &                    &                    \\
    aGrep                &0.2.1             &Tools               & \X               &                   & \X                & \Y                  &                    &                 & \X               &                    & \X                  &                  & \X               &                    &                    \\
    K-9Mail              &3.512             &Comm.       &                 &                   &                  & \Y                  &                    &                 &                 &                    &                    &                  &                 &                    &                    \\
    NetCounter           &0.14.1            &Tools               &                 &                   &                  & \Y                  &                    &                 &                 &                    &                    &                  &                 &                    & \X                  \\
    Bomber               &1                 &Casual              &                 &                   &                  & \Y                  &                    &                 &                 &                    &                    &                  &                 &                    &                    \\
    FrozenBubble         &1.12              &Puzzle              & \X               &                   &                  & \Y                  &                    &                 & \X               &                    & \X                  &                  & \X               &                    &                    \\
    AnyMemo              &8.3.1             &Education           &                 &                   & \X                & \Y                  &                    &                 &                 &                    &                    &                  & \X               & \Y                 & \X                  \\
    Blokish              &2                 &Puzzle              &                 &                   &                  & \Y                  &                    &                 &                 &                    &                    &                  &                 &                    &                    \\
    ZooBorns             &1.4.4             &Entertain.       &                 &                   &                  & \Y                  &                    &                 &                 &                    &                    &                  &                 &                    & \X                  \\
    ImportContacts       &1.1               &Tools               &                 &                   &                  & \Y                  &                    &                 &                 &                    &                    &                  &                 &                    &                    \\
    Wikipedia            &1.2.1             &Reference           &                 &                   & \X                & \Y                  &                    &                 &                 &                    &                    &                  &                 &                    &                    \\
    KeePassDroid         &1.9.8             &Tools               &                 &                   &                  & \Y                  &                    &                 &                 &                    &                    &                  &                 &                    &                    \\
    SoundBoard           &1                 &Sample              &                 &                   &                  & \Y                  &                    &                 &                 &                    &                    &                  &                 &                    &                    \\
    CountdownTimer       &1.1.0             &Tools               &                 & \Y                 &                  &                    &                    &                 &                 &                    &                    &                  &                 &                    &                    \\
    Ringdroid            &2.6               &Media               & \X               & \Y                 & \X                &                    &                    &                 & \X               &                    & \X                  &                  &                 &                    &                    \\
    SpriteMethodTest     &1.0               &Sample              &                 & \Y                 &                  &                    &                    &                 &                 &                    &                    &                  &                 &                    &                    \\
    BookCatalogue        &1.6               &Tools               &                 &                   &                  &                    &                    &                 &                 & \Y                  &                    &                  &                 &                    &                    \\
    Translate            &3.8               &Productiv.        &                 & \Y                 &                  &                    &                    &                 &                 &                    &                    &                  &                 &                    &                    \\
    TomdroidNotes        &2.0a              &Social              &                 &                   &                  &                    &                    &                 &                 & \Y                  &                    &                  &                 &                    &                    \\
    Wordpress            &0.5.0             &Productiv.        &                 &                   &                  &                    &                    &                 &                 & \Y                  &                    &                  &                 &                    & \X                  \\
    Mileage              &3.1.1             &Finance             &                 &                   &                  &                    &                    &                 &                 &                    &                    &                  &                 & \Y                  &                    \\
    Sanity               &2.11              &Comm.       &                 &                   &                  &                    &                    &                 &                 &                    &                    &                  &                 & \Y                  & \X                  \\
    DalvikExplorer       &3.4               &Tools               & \X               &                   &                  &                    &                    &                 &                 &                    &                    &                  &                 & \Y                  &                    \\
    MiniNoteViewer       &0.4               &Productiv.        &                 &                   &                  &                    &                    &                 &                 &                    &                    &                  &                 & \Y                  &                    \\
    MyExpenses           &1.6.0             &Finance             &                 &                   &                  &                    &                    &                 &                 &                    &                    &                  &                 & \Y                  & \X                  \\
    LearnMusicNotes      &1.2               &Puzzle              &                 &                   &                  &                    &                    &                 &                 &                    &                    &                  &                 & \Y                  &                    \\
    TippyTipper          &1.1.3             &Finance             &                 &                   &                  &                    &                    &                 &                 &                    &                    &                  &                 & \Y                  &                    \\
    WeightChart          &1.0.4             &Health              &                 &                   &                  &                    &                    &                 &                 &                    &                    &                  &                 & \Y                  & \X                  \\
    WhoHasMyStuff        &1.0.7             &Productiv.        &                 &                   &                  &                    &                    &                 &                 &                    &                    &                  &                 & \Y                  &      \\ \hline
  \end{tabularx}
  \egroup
  \label{tableApps}
\end{table}


\subsection{Experimental Setup}
\label{sec:protocol}

We ran our experiments on Ubuntu 14.04 virtual machines running on a
Linux server. We used \texttt{Oracle VirtualBox}\footnote{Oracle VM
  Virtualbox -- \url{http://virtualbox.org}} as our virtualization
software and \texttt{vagrant}\footnote{Vagrant --
  \url{http://vagrantup.com}} to manage these virtual machines. Each
virtual machine was configured with 2 cores and 6GB RAM. Inside the
virtual machine we installed the test input generation tools, the
Android apps, and three versions of the Android SDK versions 10
(Gingerbread), 16 (Ice-cream sandwich) and 19 (Kitkat). We chose these
versions based on their popularity, to satisfy tool dependencies and
the most recent at the time of the experiments. The emulator was
configured to use 4GB RAM and each tool was allowed to run for 1 hour
on each benchmark application.  For every run, our infrastructure
creates a new emulator with necessary tool configuration and later
destroys this emulator to avoid side-effects between tools and
applications. Given that many testing tools and applications are
non-deterministic, we repeated each experiment 10 times and we report
the mean values of all the run.

\paragraph{Coverage and System Log
  Collection}~\label{sec:coverage-collection} For each run, we
collected the code coverage for the app under test. We selected
\texttt{Emma}~\footnote{Emma: a free Java code coverage tool --
  \url{http://emma.sourceforge.net/}} as our code coverage tool
because it is officially supported and available with the Android
platform. However, since \texttt{Emma} does not allow exporting raw
statement coverage results, we parsed the HTML coverage reports to
extract line coverage information for comparison between tools. In
particular, we used this information to compute for each tool pair, 1)
the number of statements covered by both tools, and 2) the number of
statements covered by each of them separately.

To each benchmark, we added a broadcast receiver to save intermediate
coverage results to the disk. Dynodroid used a similar strategy, and
this was necessary to collect coverage from the applications before
they were restarted by the test input generation tools, and also to
track the progress of the tools at regular intervals.

SwiftHand is an exception to this protocol. This tool, in fact,
internally instruments the app under test for collecting \emph{branch}
coverage and to keep track of the app's lifecycle.  The
instrumentation is critical to the tool's functioning but conflicts
with Android's Emma instrumentation, which could not be resolved by us
within 2 weeks. Thus, we decided not to collect and compare the
statement coverage information of this tool with others. However, the
branch coverage information collected by SwiftHand on the benchmark
applications is available with our dataset.  To gather the different
failures in the app, we also collected the entire system log (also
called \texttt{logcat}), from the emulator running the app under
test. From these logs, we extracted failures that occurred while the
app was being tested in a semi-automated fashion. Specifically, we
wrote a script to find patterns of exceptions or errors in the log
file and extract them along with their available stack traces in the
log. We manually analyzed them to ignore any failures not related to
the app's execution (\eg failures of the tool themselves or
initialization errors of other apps in the Android emulator). All
unique instances of remaining failures were considered for our
results.

\vspace{-0.2cm}
\subsection{C1:  Exploration Strategy Effectiveness}
\label{sec:rq1}

Test input generation tools for Android implement different strategies
to explore as much behavior as possible of the application under
test. Section~\ref{sec:overv-andr-test} presented an overview of the
three main strategies, that is, random, model-based, and
systematic. Although some of these tools have been already evaluated
according to how much code coverage they can achieve, it is still
unclear whether there is any strategy that is better than others in
practice. Previous evaluations were either incomplete because they did
not include comparison with other tools, or they were, according to
our opinion, biased. Since we believe that the most critical resource
is \emph{time}, tools should evaluate how much coverage they can
achieve within a certain time limit. Tools such as Dynodroid and
EvoDroid, instead, have been compared to Monkey by comparing the
coverage achieved given the same number of generated events.  We thus
set up the experiment by running each tool 10 times on each
application of Table~\ref{tableApps}, and we collected the achieved
coverage as described in Section~\ref{sec:protocol}.

\begin{figure}[t]
  \centering
  \includegraphics[trim = 10mm 20mm 10mm 10mm, width=\linewidth]{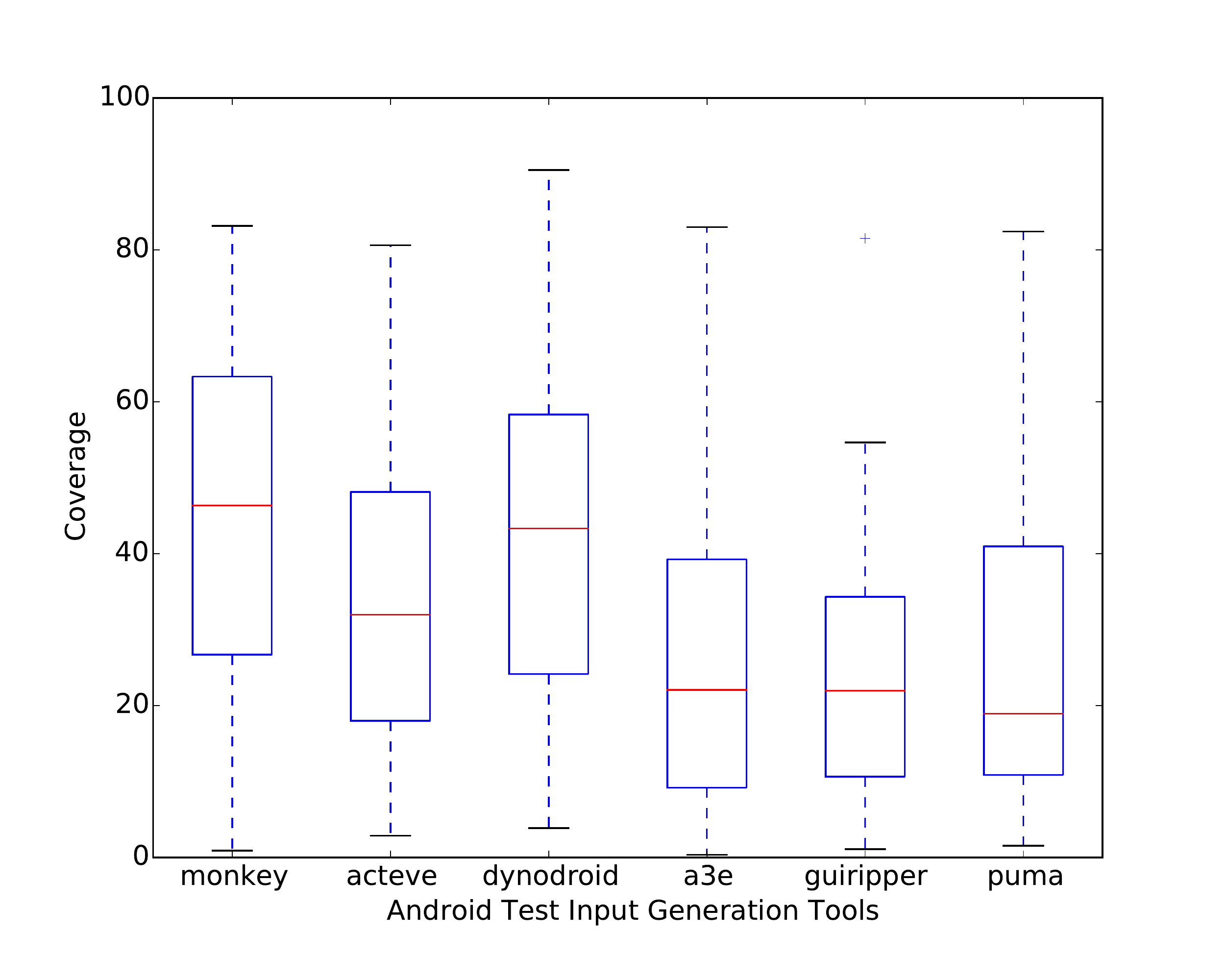}
  \caption{Variance of the coverage across the applications over 10
    runs.}
  \label{fig:variance}
\end{figure}

Figure~\ref{fig:variance} reports the variance of the mean coverage of
10 runs that each tool achieved on the considered benchmarks. We can see
that on average Dynodroid and Monkey perform better than other tools,
followed by \ACTEVE. The other three tools (\ie \AEEE, GUIRipper and
PUMA) achieve quite a low coverage.

Despite this, even those tools that on average achieve low coverage
can reach very high coverage (approximately 80\%) for few apps. We
manually investigated these apps, and we saw that these are, as
expected, the most simple ones. The two outliers for which every tool
achieved very high coverage are DivideAndConquer and
RandomMusicPlayer. The first one is a game that accepts only touches
and swipes as events, and they can be provided without much logic in
order to proceed with the game. RandomMusicPlayer is a music player
that randomly plays music. The possible user actions are quite
limited, as there is only one activity with 4 buttons.

Similarly, there are some applications for which every tool, even the
ones that performed best, achieved very low coverage (\ie lower than
5\%). Two of these apps, K9mail (a mail client) and PasswordMakerPro
(an app to generate and store authentication data), highly depend on
external factors, such as the availability of a valid account. Such
inputs are nearly impossible to generate automatically, and therefore
every tool stalls at the beginning of the exploration.
Few tools provide an option to manually interact with application at first, and
then use the tool to perform subsequence test input generation. However, we did
not use this feature for scalability reasons and concerns of giving such tools
an unfair advantage of the manual intelligence.


\begin{figure}[t]
  \centering
  \includegraphics[trim = 10mm 15mm 10mm 10mm, width=\columnwidth]{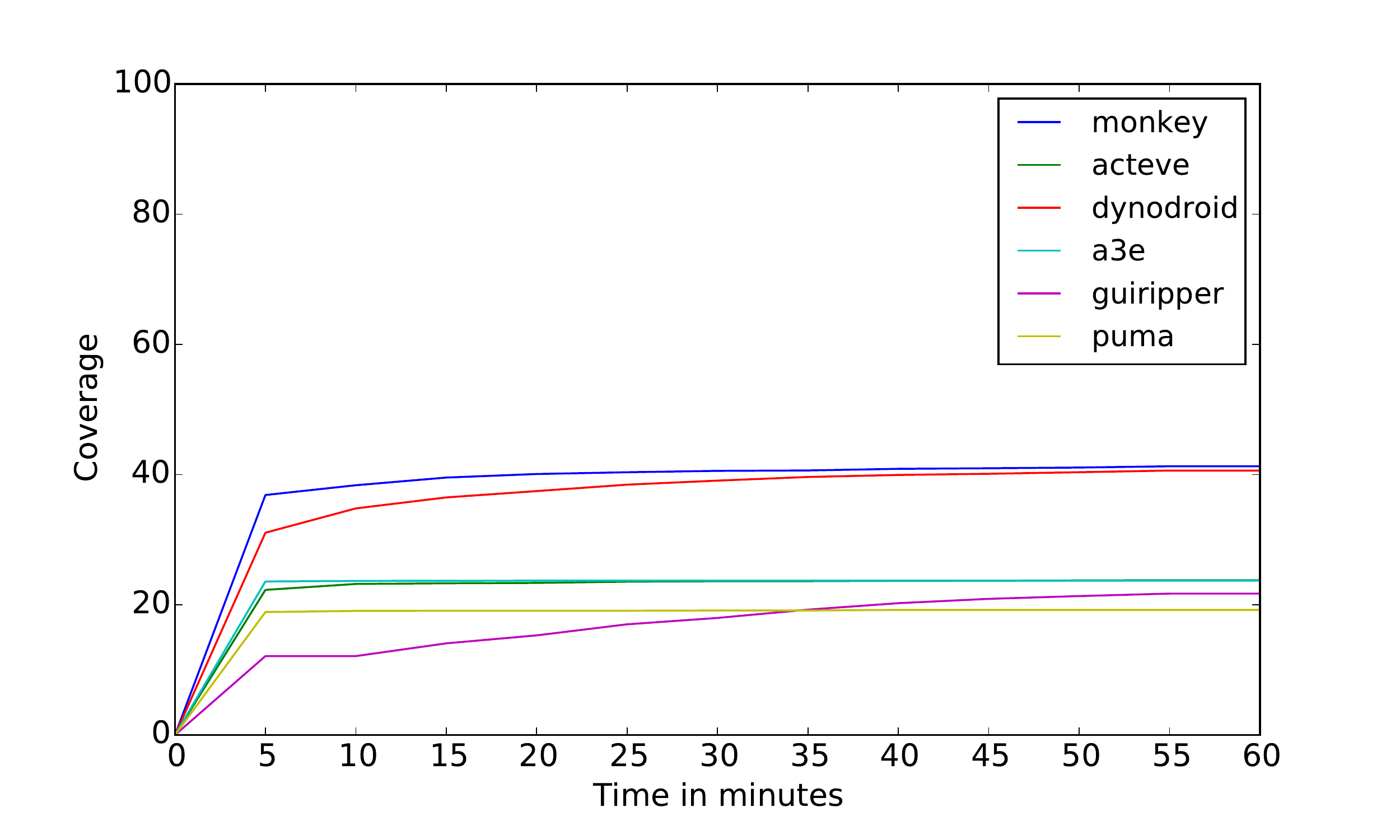}
  \caption{Progressive coverage}
  \label{fig:progress}
\end{figure}

Figure~\ref{fig:progress} reports the progressive coverage of each
tool over the maximum time bound we gave, \ie 60 minutes. The plot
reports the mean coverage across all apps over the 10 runs. This plot
evidences an interesting finding, that is, all tools can hit their
maximum coverage within few minutes (between 5 and 10 minutes). The
only exception to this is GUIRipper. The likely reason why this
happens in that GUIRipper frequently restarts the exploration from the
starting state, and this operation takes time. This is the main
problem that SwiftHand addresses by implementing an exploration
strategy that limits the number of restarts.

\subsection{C2: Fault Detection Ability}
\label{sec:rq2}

\begin{figure}[t]
  \centering
  \includegraphics[trim = 10mm 10mm 0mm 0mm, width=\linewidth]{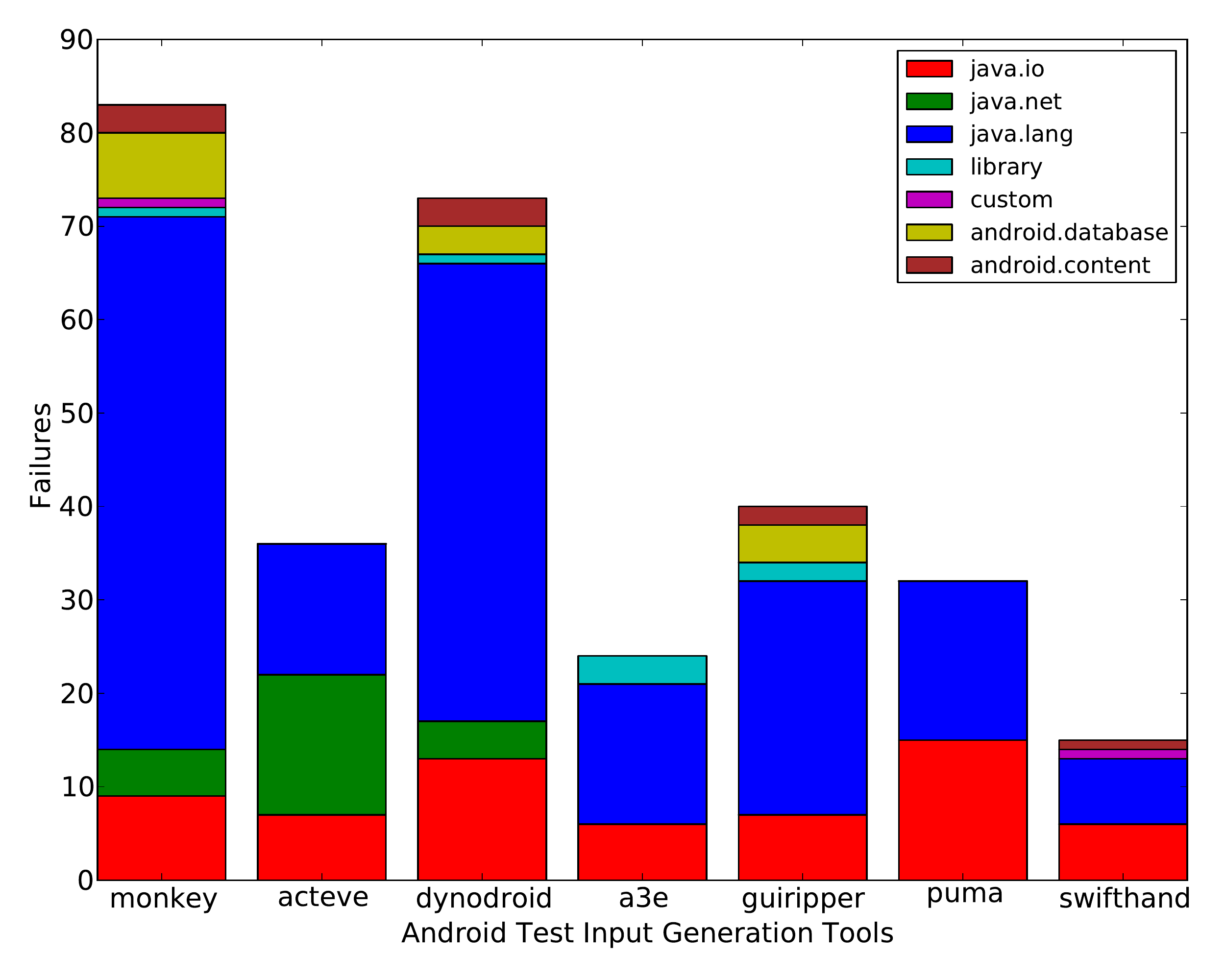}
  \caption{Distribution of the failures triggered by the tools.}
  \vspace{-16pt}
  \label{fig:failures}
\end{figure}

The final goal of test input generation tools is to expose faults in
the app under test. Therefore, beside code coverage, we checked how
many failures each tool can reveal with a time budget of one hour per
app. None of the Android tools can identify failures other than runtime
exceptions, although there is some promising work that goes in that
direction~\cite{Zaeem:OraclesAndroid:ICST:2014}.

Figure~\ref{fig:failures} reports the results of this study. Numbers
on the y axis represent the \emph{cumulative unique failures} across
the 10 runs across all apps. We consider a failure unique when its
stack trace differs from other ones.

The plot also reports some high level statistics about which are the
most frequent failures (we report the package name of the runtime
exception). Only few of them are custom exceptions (\ie exceptions
that are declared in the app under test). The vast majority of them
generate standard Java exceptions, and among them the most frequent
ones are Null Pointer Exceptions.

Because SwiftHand is the tool with the worst performance in this part
of the evaluation, we looked into its results in more detail to
understand the reasons behind that. We found that SwiftHand crashes on
many of our benchmarks, which prevents it from producing useful
results in these cases. Further analysis revealed that the crashes are
most likely due to SwiftHand's use of \texttt{asmdex} to instrument
the apps it is testing. The \texttt{asmdex} framework is in fact known
to be buggy, not well maintained, and crash-prone.

\begin{figure*}[!th]
  \centering
  \includegraphics[trim = 0mm 10mm 55mm 10mm, width=1\linewidth]{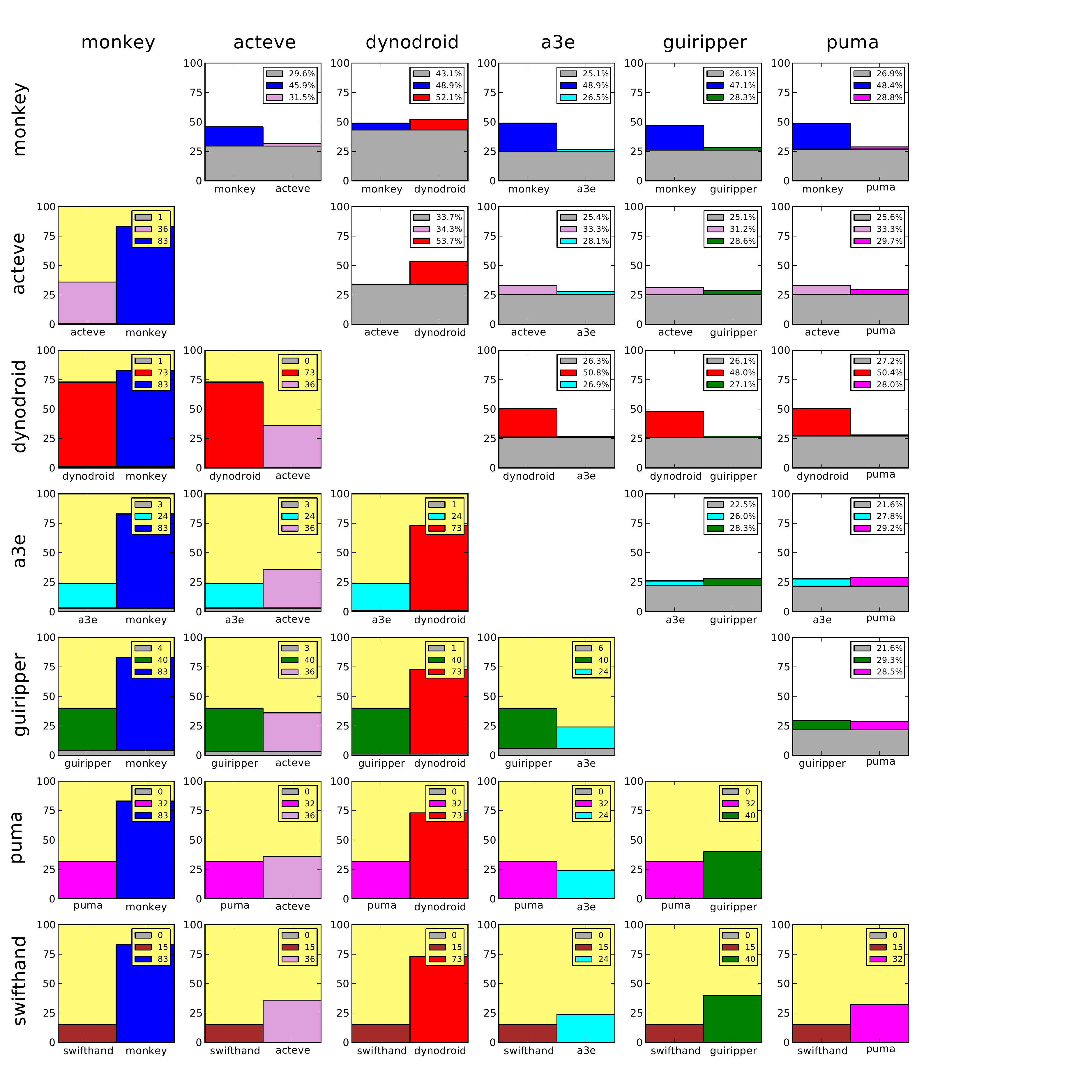}
  \caption{Pairwise comparison of tools in terms of coverage and failures
  triggered. The plots on top-right show percent statement coverage of the tools
  and the ones in the bottom-left section show absolute number of failures
  triggered. The gray bars in all plots show commonalities between the pairs of
  tools .}
  \label{fig:pairwise}
\end{figure*}

Figure~\ref{fig:pairwise} reports a pairwise comparison of each tool
according to coverage (upper part of the Figure, \ie boxes with white
background) and fault detection ability (lower part of the Figure, \ie
boxes with yellow background). In this figure, we compare the coverage
of the best run out of 10 runs for each tool, whereas the number of
failures reported are the cumulative failures across 10 runs with the
same stack trace. The comparison reports which lines are covered (and
respectively which failures are reported) by both tools (reported in
grey), and which lines are covered by only one of the two
tools. Results show that tools do not complement each other in terms
of code coverage, but they do in terms of fault detection. In fact,
while for coverage the common parts are significant, it is almost the
opposite in terms of failures.

\subsection{C3: Ease of Use}
\label{sec:rq3}

Tools should ideally work out of the box, and should not require extra
effort of the user in terms of
configuration. Table~\ref{tab:ease-compatibility} reports whether the
tool worked out of the box (NO\_EFFORT), whether it required some
effort (LITTLE\_EFFORT), either to properly configure it or to fix
minor issues, or whether it required major efforts (MAJOR\_EFFORT) to
make it work.


\begin{table}
  \centering
  \caption{Ease of use and compatibility of each tool with the most
    common Android framework versions.}
  \begin{tabular}[h]{|l|c|c|}
    \hline
    Name &
    Ease Use &
    Compatibility
    \\
    \hline
    Monkey~\cite{monkey} & NO\_EFFORT & any\\
    Dynodroid~\cite{machiry13fse} & NO\_EFFORT & v.2.3 \\
    %
    GUIRipper~\cite{amalfitano12ase} & MAJOR\_EFFORT & any \\
    \AEEE-Depth-first~\cite{azim13oopsla} & LITTLE\_EFFORT & any \\
    SwiftHand~~\cite{choi13oopsla} & MAJOR\_EFFORT & v.4.1+ \\
    PUMA~\cite{Hao:PUMA:Mobisys:14} & LITTLE\_EFFORT & v.4.3+ \\
    ACTEve~\cite{anand12fse} & MAJOR\_EFFORT & v.2.3 \\
    \hline
  \end{tabular}
  \label{tab:ease-compatibility}
\end{table}

We now briefly report our experience with installing each tool, and we
describe the required fixes to make each of them work. Some of the
 changes were required to make the tools run on our infrastructure.

\textbf{Monkey:} We used the vanilla Monkey from the Android
distribution for our experimentation. The monkey tool was configured
to ignore any crash, system timeout or security exceptions during the
experiment and to continue till the experiment timeout was reached. In
addition, we configured it to wait 200 milliseconds between actions,
as this same delay value was also used in other tools. Configuring
Monkey for our infrastructure required no extra effort.

\textbf{\ACTEVE:} We consulted the authors to apply minor fixes
to the instrumentation component of \ACTEVE, which instruments
both the Android SDK and the app under test. While generating tests
\ACTEVE often restarts the application. To ensure that we do not
lose the coverage information, we modified \ACTEVE to save the
intermediate coverage before application restarts.

\textbf{GUIRipper:} GUIRipper is available as a binary
distribution with batch scripts to run it in the Windows OS. We
reversed engineered GUIRipper and wrote shell scripts to port it
to our Linux based infrastructure. We configured the tool to use its
systematic ripping strategy instead of the default random
strategy. During the experiments, GUIRipper often restarts the
Android emulator from a snapshot to return to the initial state. We
modified the tool to save intermediate coverage before such restarts.

\textbf{Dynodroid:} We obtained a running version of Dynodroid from
the virtual machine provided on the tool's page. Dynodroid is tightly
coupled with the Android emulator, for which the tool includes an
instrumented system image. In addition, the tool performs an extensive
setup of the device after boot before starting the exploration. We
configured our timer to start counting at the start of the actual
exploration to account for this one time setup time.

\textbf{\AEEE:} We updated \AEEE's dependencies to make it work with
the latest version of the Android SDK. The public release only
supports the depth-first exploration strategy for systematic testing,
which was used in our experiments. In addition, we modified the tool
to report verbose results for Android commands invoked, and to not
shutdown the emulator after input generation to allow us to collect
reports from it.

\textbf{SwiftHand:} The SwiftHand tool consists of two components:
front-end, which performs bytecode instrumentation of the apps, and
back-end, which performs test input generation. We fixed the
dependencies of the front-end tool by obtaining an older version of
the \texttt{asmdex} library ({\small
  \url{http://asm.ow2.org/asmdex-index.html}}), and wrote a wrapper
shell script to connect the two components in our infrastructure.

\textbf{PUMA:} To get PUMA running, we applied a minor patch to use an
alternate API for taking device screenshots. We also altered the different
timeouts in the tool to match our experimental settings.

\subsection{C4: Android Framework Compatibility}
\label{sec:rq4}

Android developers have to constantly deal with the fragmentation
problem, that is, their application has to run on devices that have
different hardware characteristics and use different versions of the
Android framework. It is therefore desirable of a test input generator
tool to be compatible with multiple releases of the Android framework
in order to let developers assess the quality of their app in
different environments.
Therefore, we ran each tool on three popular Android framework
releases, as described in Section~\ref{sec:protocol}, and assessed
whether it could work correctly.

Table~\ref{tab:ease-compatibility} reports the results of this
study. The table shows that 4 out of 7 tools do not offer this
feature. Some tools (PUMA and SwiftHand) are compatible only with the
most recent releases of the Android Framework, while others (\ACTEVE
and Dynodroid) are bound to a very old one. \ACTEVE and Dynodroid
could be compatible with other frameworks, but this would require to
instrument them first. SwiftHand and PUMA, instead, are not compatible
with older releases of the Android Framework because they use features
of the underlying framework that are not available in previous
releases.

\section{Discussion and Future Research Directions}
\label{sec:discussion}

The experiments presented in Section~\ref{sec:evaluation} report
unexpected results: the random exploration strategies implemented by
Monkey and Dynodroid can obtain higher coverage than more
sophisticated strategies implemented by other tools.
It thus seems that Android applications are different from Java
stand-alone application, where random strategies have been shown to be
highly inefficient compared to systematic
strategies~\cite{Memon:GUITAR:WCRE:2003,Mariani:Autoblacktest:ICST:2012,Gross:Exsyst:ICSE:2012}. Our
results show that 1) most of the behavior can be exercised by
generating only UI events, and 2) to expose this behavior the random
approach is effective enough.

Considering the four criterion of the study, Monkey would clearly be
the winner among the existing test input generation tools, since it
achieves, on average, the best coverage, it can report the largest
number of failures, it is ease to use, and works for every platform.
This does not mean that the other tools should not be
considered. Actually it is quite the opposite, since every other tool
has strong points that, if properly implemented and properly combined
can lead to significant improvements. We now list some of the features
that some tools already implement and should be considered by other
tools:

\begin{description}
\addtolength{\itemsep}{-2mm}
\item[System events.] Dynodroid and \ACTEVE can generate system events
  beside standard UI events. Even if the behavior of an app may depend
  only partially on system events, generating them can reveal failures
  that would be hard to uncover otherwise.

\item[Minimize restarts.] Progressive coverage shows that tools that
  frequently restart their exploration from the starting point need
  more time to achieve their maximum coverage. The search algorithm
  that Swifthand implements aims to minimize such restarts, and thus
  allows tools to achieve high coverage in less time.

\item[Manually provided inputs.] Specific behaviors can sometimes only
  be explored by providing specific inputs, which may be hard to
  generate randomly or by means of systematic techniques. Some tools,
  such as Dynodroid and GUIRipper let users manually provide values
  that the tool can later use during the analysis. This feature is
  highly desirable, since it allows tools to explore the app in
  presence of login forms and similar complex inputs.

\item[Multiple starting states.] The behavior of many applications
  depend on the underlying content providers. An email client, for
  instance, would show an empty inbox, unless the content provider
  contains some messages. GUIRipper starts exploring the application
  from different starting states (\eg when the content provider is
  empty and when it has some entries). Even if this has to be done
  manually by the user, by properly creating snapshots of the app, it
  allows to potentially explore behavior that would be hard to explore
  otherwise.

\item[Avoid side effects among different runs.] Using the tool on
  multiple apps requires to reset the environment to avoid
  side-effects across multiple runs. Although, in our experiments we 
  used a fresh emulator instance between runs, we realized that some 
  tools, such as Dynodroid and \AEEE, had capabilities to (partially) 
  clean up the environment by uninstalling the application and deleting 
  its data. We believe that every tool should reuse the environment 
  for efficiency reasons but should do it without affecting its accuracy.
\end{description}

During our study, we also identified limitations that significantly
affect the effectiveness of all tools. We report these limitations,
together with desirable and missing features, such that they could be
the focus of future research in this area:
\begin{description}
\addtolength{\itemsep}{-2mm}
\item[Reproducible test cases.] None of the tools allows to easily
  reproduce failures. They report uncaught runtime exceptions on the
  logfile, but they do not generate test cases that can be later
  rerun. We believe that this is an essential feature that every tool
  of this type should have.

\item[Debugging support.] The lack of reproducible test cases makes it
  hard to identify the root cause of the failures. The stack trace of
  the runtime exception is the only information that a developer can
  use, and this information is lost in the execution logs. Testing
  tools for Android should make failures more visible, and should
  provide more information to ease debugging. In our evaluation, we had
  a hard time understanding if failures were caused by real faults or
  rather were caused by limitations of the emulator. More information
  about each failure would have helped in this task.

\item[Mocking.] Most apps for which tools had low coverage highly
  depend on environment elements such as content providers. It is
  impossible to explore most of K9mail functionalities unless there is
  a valid email account already set up and unless there are existing
  emails in the account. GUIRipper alleviates this problem by letting
  users prepare different snapshots of the app. We believe that
  working on a proper mocking infrastructure for Android apps would be
  a significant contribution, as it would lead to drastic code
  coverage improvements.

\item[Sandboxing.] Testing tools should also provide proper sandboxing,
  that is, they should block operations that may potentially have
  disruptive effects (for instance sending emails using a valid
  account, or allow critical changes using a real social networking
  account). None of the tools keeps this problem into account.

\item[Focus of fragmentation problem.] While C4 of our evaluation
  showed that some tools can run on multiple versions of the Android
  framework, none of them is specifically designed for cross-device
  testing. While this is a different testing problem, we believe that
  testing tools for Android should also move towards this direction,
  as fragmentation is the major problem that Android developers have
  to face.
\end{description}

\section{Conclusion}
\label{sec:conclusion}

In this paper, we presented a comparative study of the main existing
test input generation techniques and corresponding tools for Android.
We evaluated these tools according to four criteria: code coverage,
fault detection capabilities, ease of use, and compatibility with
multiple Android framework versions.  After presenting the results of
this comparison, we discuss strengths and weaknesses of the different
techniques and highlight potential venues for future research in this
area. All of our experimental infrastructure and data are publicly
available at {\small
  \url{http://www.cc.gatech.edu/~orso/software/androtest}}, so that
other researchers can replicate our studies and build on our work.

\section*{Acknowledgments}

We would like to thank the authors of the tools, (specifically, Saswat
Anand, Domenico Amalfitano, Aravind Machiry, Tanzirul Azim, Wontae
Choi, and Shuai Hao) for making their tools available and for
answering our clarification questions regarding the tool setup.

\balance

\bibliographystyle{abbrv}
\bibliography{paper}

\begin{thebibliography}{10}

\bibitem{amalfitano12ase}
D.~Amalfitano, A.~R. Fasolino, P.~Tramontana, S.~De~Carmine, and A.~M. Memon.
\newblock {Using GUI Ripping for Automated Testing of Android Applications}.
\newblock In {\em Proceedings of the 27th IEEE/ACM International Conference on
  Automated Software Engineering}, ASE 2012, pages 258--261, New York, NY, USA,
  2012. ACM.

\bibitem{MobiGUITARIEEESoftware2014}
D.~Amalfitano, A.~R. Fasolino, P.~Tramontana, B.~D. Ta, and A.~M. Memon.
\newblock {MobiGUITAR} -- a tool for automated model-based testing of mobile
  apps.
\newblock {\em IEEE Software}, PP(99):NN--NN, 2014.

\bibitem{anand12fse}
S.~Anand, M.~Naik, M.~J. Harrold, and H.~Yang.
\newblock {Automated Concolic Testing of Smartphone Apps}.
\newblock In {\em Proceedings of the ACM SIGSOFT 20th International Symposium
  on the Foundations of Software Engineering}, FSE '12, pages 59:1--59:11, New
  York, NY, USA, 2012. ACM.

\bibitem{FlowDroid}
S.~Arzt, S.~Rasthofer, C.~Fritz, E.~Bodden, A.~Bartel, J.~Klein, Y.~Le~Traon,
  D.~Octeau, and P.~McDaniel.
\newblock {FlowDroid}: Precise context, flow, field, object-sensitive and
  lifecycle-aware taint analysis for {Android} apps.
\newblock In {\em Proceedings of the 35th ACM SIGPLAN Conference on Programming
  Language Design and Implementation}, PLDI '14, pages 259--269, New York, NY,
  USA, 2014. ACM.

\bibitem{azim13oopsla}
T.~Azim and I.~Neamtiu.
\newblock {Targeted and Depth-first Exploration for Systematic Testing of
  Android Apps}.
\newblock In {\em Proceedings of the 2013 ACM SIGPLAN International Conference
  on Object Oriented Programming Systems Languages \& Applications}, OOPSLA
  '13, pages 641--660, New York, NY, USA, 2013. ACM.

\bibitem{Bartel:Dexpler:SOAP:2012}
A.~Bartel, J.~Klein, Y.~Le~Traon, and M.~Monperrus.
\newblock {Dexpler: Converting Android Dalvik Bytecode to Jimple for Static
  Analysis with Soot}.
\newblock In {\em Proceedings of the ACM SIGPLAN International Workshop on
  State of the Art in Java Program Analysis}, SOAP '12, pages 27--38, New York,
  NY, USA, 2012. ACM.

\bibitem{choi13oopsla}
W.~Choi, G.~Necula, and K.~Sen.
\newblock {Guided GUI Testing of Android Apps with Minimal Restart and
  Approximate Learning}.
\newblock In {\em Proceedings of the 2013 ACM SIGPLAN International Conference
  on Object Oriented Programming Systems Languages \& Applications}, OOPSLA
  '13, pages 623--640, New York, NY, USA, 2013. ACM.

\bibitem{webmate-swqd-2013}
V.~Dallmeier, M.~Burger, T.~Orth, and A.~Zeller.
\newblock {WebMate: Generating Test Cases for Web 2.0}.
\newblock In {\em Software Quality. Increasing Value in Software and Systems
  Development}, pages 55--69. Springer, 2013.

\bibitem{Enck:Taintdroid:OSDI:2010}
W.~Enck, P.~Gilbert, B.-G. Chun, L.~P. Cox, J.~Jung, P.~McDaniel, and A.~N.
  Sheth.
\newblock {TaintDroid}: An information-flow tracking system for realtime
  privacy monitoring on smartphones.
\newblock In {\em Proceedings of the 9th USENIX Conference on Operating Systems
  Design and Implementation}, OSDI'10, pages 1--6, Berkeley, CA, USA, 2010.
  USENIX Association.

\bibitem{Gorla:CHABADA:ICSE:2014}
A.~Gorla, I.~Tavecchia, F.~Gross, and A.~Zeller.
\newblock Checking app behavior against app descriptions.
\newblock In {\em Proceedings of the 36th International Conference on Software
  Engineering}, ICSE 2014, pages 1025--1035, New York, NY, USA, June 2014. ACM.

\bibitem{Gross:Exsyst:ICSE:2012}
F.~Gross, G.~Fraser, and A.~Zeller.
\newblock {EXSYST: Search-based GUI Testing}.
\newblock In {\em Proceedings of the 34th International Conference on Software
  Engineering}, ICSE '12, pages 1423--1426, Piscataway, NJ, USA, 2012. IEEE
  Press.

\bibitem{Hao:PUMA:Mobisys:14}
S.~Hao, B.~Liu, S.~Nath, W.~G. Halfond, and R.~Govindan.
\newblock {PUMA}: Programmable {UI}-automation for large-scale dynamic analysis
  of mobile apps.
\newblock In {\em Proceedings of the 12th Annual International Conference on
  Mobile Systems, Applications, and Services}, MobiSys '14, pages 204--217, New
  York, NY, USA, 2014. ACM.

\bibitem{Hu:GUITestingAndroid:AST:2011}
C.~Hu and I.~Neamtiu.
\newblock {Automating GUI Testing for Android Applications}.
\newblock In {\em Proceedings of the 6th International Workshop on Automation
  of Software Test}, AST '11, pages 77--83, New York, NY, USA, 2011. ACM.

\bibitem{Kechagia:AndroidFailures:EMSE:2014}
M.~Kechagia, D.~Mitropoulos, and D.~Spinellis.
\newblock Charting the {API} minefield using software telemetry data.
\newblock {\em Empirical Software Engineering}, pages 1--46, 2014.

\bibitem{LiuJPF-Android}
Y.~Liu, C.~Xu, and S.~Cheung.
\newblock Verifying android applications using java pathfinder.
\newblock Technical report, The Hong Kong University of Science and Technology,
  2012.

\bibitem{google-bouncer}
H.~Lockheimer.
\newblock Google bouncer.
\newblock
  \url{http://googlemobile.blogspot.com.es/2012/02/android-and-security.html}.

\bibitem{machiry13fse}
A.~Machiry, R.~Tahiliani, and M.~Naik.
\newblock {Dynodroid: An Input Generation System for Android Apps}.
\newblock In {\em Proceedings of the 2013 9th Joint Meeting on Foundations of
  Software Engineering}, ESEC/FSE 2013, pages 224--234, New York, NY, USA,
  2013. ACM.

\bibitem{mahmood14fse}
R.~Mahmood, N.~Mirzaei, and S.~Malek.
\newblock {EvoDroid: Segmented Evolutionary Testing of Android Apps}.
\newblock In {\em Proceedings of the 22nd ACM SIGSOFT International Symposium
  on Foundations of Software Engineering}, FSE 2014, New York, NY, USA, 2014.
  ACM.

\bibitem{Maji:robustness:DSN:2012}
A.~K. Maji, F.~A. Arshad, S.~Bagchi, and J.~S. Rellermeyer.
\newblock An empirical study of the robustness of inter-component communication
  in android.
\newblock In {\em Proceedings of the 2012 42Nd Annual IEEE/IFIP International
  Conference on Dependable Systems and Networks (DSN)}, DSN '12, pages 1--12,
  2012.

\bibitem{Mariani:Autoblacktest:ICST:2012}
L.~Mariani, M.~Pezze, O.~Riganelli, and M.~Santoro.
\newblock {AutoBlackTest: Automatic Black-Box Testing of Interactive
  Applications}.
\newblock In {\em Proceedings of the 2012 IEEE Fifth International Conference
  on Software Testing, Verification and Validation}, ICST '12, pages 81--90,
  Washington, DC, USA, 2012. IEEE Computer Society.

\bibitem{Memon:GUITAR:WCRE:2003}
A.~Memon, I.~Banerjee, and A.~Nagarajan.
\newblock {GUI Ripping: Reverse Engineering of Graphical User Interfaces for
  Testing}.
\newblock In {\em Proceedings of the 10th Working Conference on Reverse
  Engineering}, WCRE '03, pages 260--, Washington, DC, USA, 2003. IEEE Computer
  Society.

\bibitem{crawljax:tweb12}
A.~Mesbah, A.~van Deursen, and S.~Lenselink.
\newblock {Crawling {Ajax}-based Web Applications through Dynamic Analysis of
  User Interface State Changes}.
\newblock {\em ACM Transactions on the Web (TWEB)}, 6(1):3:1--3:30, 2012.

\bibitem{monkey}
The {Monkey UI} android testing tool.
\newblock \url{http://developer.android.com/tools/help/monkey.html}.

\bibitem{nullfuzzer}
Intent fuzzer, 2009.
\newblock
  \url{http://www.isecpartners.com/tools/mobile-security/intent-fuzzer.aspx}.

\bibitem{Octeau:Dare:FSE:2012}
D.~Octeau, S.~Jha, and P.~McDaniel.
\newblock {Retargeting Android Applications to Java Bytecode}.
\newblock In {\em Proceedings of the ACM SIGSOFT 20th International Symposium
  on the Foundations of Software Engineering}, FSE '12, pages 6:1--6:11, New
  York, NY, USA, 2012. ACM.

\bibitem{mobileVsDesktop}
{Rebecca Murtagh}.
\newblock {Number of apps available in leading app stores as of July 2014}.
\newblock
  http://searchenginewatch.com/article/2353616/Mobile-Now-Exceeds-PC-The-Biggest-Shift-Since-the-Internet-Began,
  July 2014.

\bibitem{RoyChoudhary:XPERT:ICSE:2013}
S.~Roy~Choudhary, M.~R. Prasad, and A.~Orso.
\newblock {X-PERT: Accurate Identification of Cross-browser Issues in Web
  Applications}.
\newblock In {\em Proceedings of the 2013 International Conference on Software
  Engineering}, ICSE '13, pages 702--711, Piscataway, NJ, USA, 2013. IEEE
  Press.

\bibitem{sasnauskas14woda}
R.~Sasnauskas and J.~Regehr.
\newblock {Intent Fuzzer: Crafting Intents of Death}.
\newblock In {\em Proceedings of the 2014 Joint International Workshop on
  Dynamic Analysis (WODA) and Software and System Performance Testing,
  Debugging, and Analytics (PERTEA)}, WODA+PERTEA 2014, pages 1--5, New York,
  NY, USA, 2014. ACM.

\bibitem{smali}
{Smali/baksmali, an assembler/disassembler for the dex format used by Dalvik}.
\newblock \url{https://code.google.com/p/smali}.

\bibitem{appStoreNumbers}
{Statista}.
\newblock {Number of apps available in leading app stores as of July 2014}.
\newblock
  \url{http://www.statista.com/statistics/276623/number-of-apps-available-in-leading-app-stores/},
  August 2014.

\bibitem{vanderMerwe12sigsoft}
H.~van~der Merwe, B.~van~der Merwe, and W.~Visser.
\newblock {Verifying Android Applications Using Java PathFinder}.
\newblock {\em SIGSOFT Softw. Eng. Notes}, 37(6):1--5, Nov. 2012.

\bibitem{vanderMerwe14sigsoft}
H.~van~der Merwe, B.~van~der Merwe, and W.~Visser.
\newblock {Execution and Property Specifications for JPF-android}.
\newblock {\em SIGSOFT Softw. Eng. Notes}, 39(1):1--5, Feb. 2014.

\bibitem{Wei:ProfileDroid:mobicom:2012}
X.~Wei, L.~Gomez, I.~Neamtiu, and M.~Faloutsos.
\newblock {ProfileDroid: Multi-layer Profiling of Android Applications}.
\newblock In {\em Proceedings of the 18th Annual International Conference on
  Mobile Computing and Networking}, Mobicom '12, pages 137--148, New York, NY,
  USA, 2012. ACM.

\bibitem{yang13fase}
W.~Yang, M.~R. Prasad, and T.~Xie.
\newblock {A Grey-box Approach for Automated GUI-model Generation of Mobile
  Applications}.
\newblock In {\em Proceedings of the 16th International Conference on
  Fundamental Approaches to Software Engineering}, FASE'13, pages 250--265,
  Berlin, Heidelberg, 2013. Springer-Verlag.

\bibitem{ye13momm}
H.~Ye, S.~Cheng, L.~Zhang, and F.~Jiang.
\newblock {DroidFuzzer: Fuzzing the Android Apps with Intent-Filter Tag}.
\newblock In {\em Proceedings of International Conference on Advances in Mobile
  Computing \&\#38; Multimedia}, MoMM '13, pages 68:68--68:74, New York, NY,
  USA, 2013. ACM.

\bibitem{Zaeem:OraclesAndroid:ICST:2014}
R.~N. Zaeem, M.~R. Prasad, and S.~Khurshid.
\newblock Automated generation of oracles for testing user-interaction features
  of mobile apps.
\newblock In {\em Proceedings of the 2014 IEEE International Conference on
  Software Testing, Verification, and Validation}, ICST '14, pages 183--192,
  Washington, DC, USA, 2014. IEEE Computer Society.

\end{thebibliography}

\end{document}